\documentclass[a4paper]{jpconf}
\usepackage{graphicx}
\begin{document}
\title{ Study of charge-phase diagrams for coupled system of Josephson junctions}

\author {M. Hamdipour$^{1,2}$   and  Yu. M. Shukrinov$^{1}$}

\address{$^1$  BLTP, JINR, Dubna, Moscow Region, 141980, Russia}
\address{$^2$  Department of Physics, Institute for Advanced Studies in Basic Sciences, P.O.Box 45195-1159, Zanjan, Iran}

\ead{hamdipur@theor.jinr.ru}

\begin{abstract}
Dynamics of stacked  intrinsic Josephson junctions (IJJ) in the high-Tc superconductors is theoretically
investigated. We calculate the current-voltage characteristics (CVC) of IJJ   and study the breakpoint region on
the outermost branch of the CVC for the stacks with 9 IJJ. A method for investigation of the fine structure in
CVC of IJJ based on the recording the "phase-charge" diagrams is suggested.  It is demonstrated that this method
reflects the main features of the breakpoint region.
\end{abstract}

\section{Introduction and model}
Study of intrinsic Josephson junctions in HTSc, like $Bi_2Sr_2CaCu_2O_{8+\delta}$ shows interesting physical
features. In Refs.\cite{sm-sust07,sm-prl07,smp-prb07} we studied the multiple branch structure of the CVC of IJJ
and showed that the branches have a breakpoint (BP) and some breakpoint region (BPR) before transition to the
another branch. The BP  is determined by the creation of the longitudinal plasma wave (LPW) with a definite wave
number $k$, which depends on the coupling parameter $\alpha$, dissipation parameter $\beta$, the number of
junctions in the stack, and the boundary conditions.

Here we show that using the phase-charge diagrams (phase portraits) give us an additional method for
investigation of the fine structure in the CVC. We use the $(CCJJ+DC)$ - model \cite{sms-physC06} to investigate
the phase dynamics and CVC of coupled system of Josephson junctions. The system of equations in this model has a
form
\begin{eqnarray}
\frac{d^2}{dt^2}\varphi_{l}=(I-\sin \varphi_{l} -\beta\frac{d\varphi_{l}}{dt})+ \alpha (\sin \varphi_{l+1}+
\sin\varphi_{l-1} \nonumber \\- 2\sin\varphi_{l})+ \alpha
\beta(\frac{d\varphi_{l+1}}{dt}+\frac{d\varphi_{l-1}}{dt}-2\frac{d\varphi_{l}}{dt}) \label{d-phi-dif}
\end{eqnarray}
Here $\varphi_l$ is the gauge-invariant phase differences $\varphi_l(t)=
\theta_{l+1}(t)-\theta_{l}(t)-\frac{2e}{\hbar}\int^{l+1}_{l}dz A_{z}(z,t)$  between superconducting layers
($S$-layers), $\theta_{l}$ is the phase of the order parameter in S-layer $l$, $A_z$ is the vector potential in
the barrier.  The current, voltage and time are normalized to the critical current $I_c$, $V_0=\frac{\hbar
\omega_p}{2 e}$ and inverse of plasma frequency $\omega_{p}^{-1}$, respectively. The details of simulation are
available at Refs. \cite{koyama96,smp-prb07,shk-prb09}. We  calculate the charge on S-layer as
$Q_l(t)=\alpha(V_l(t)-V_{l-1}(t))$, where $V_l(t)$ is the voltage drop on insulator layer l at time moment $t$.
We  start from the current value I=1.2 and decrease it up to the transition to another branch. We record the
time dependence of the phase difference between layers $l$ and $l-1$ and the electric charge on the layer $l$ at
some selective points of the BPR.

\begin{figure}[h]
\centering
\includegraphics[height=90mm]{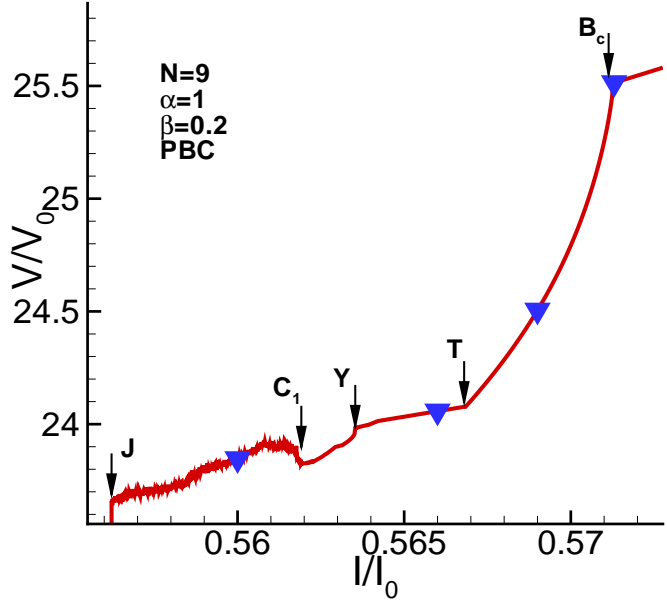}
\caption{\label{iv}(Color online) The BPR part of the outermost branch for the stack with 9 IJJ at $\alpha=1$,
$\beta=0.2$ and periodic boundary conditions. The symbol $\nabla$ indicates the points at which the time
dependence of the phase difference and the charge on the layer are investigated. }
\end{figure}

\section{Phase-charge diagrams}

The BPR part of the outermost branch of CVC for the stack with 9 IJJ at $\alpha=1$ and $\beta=0.2$ and periodic
boundary conditions is presented in Fig.~\ref{iv}. It indicates the current values, at which we have
investigated the phase-charge diagrams.  The phase-charge diagram is a trajectory of junction in its phase
space:($\varphi,Q$). We suggest to use this diagrams as a tool to probe the dynamics of system. For periodic
systems like simple harmonic oscillator, the phase portraits  in ($\varphi$,$\dot{\varphi}$) space are circles
with their centers at the origin of coordinate system. For the system of IJJ it is natural to investigate the
phase-charge diagrams.

\begin{figure}[h]
\centering
\includegraphics[height=90mm]{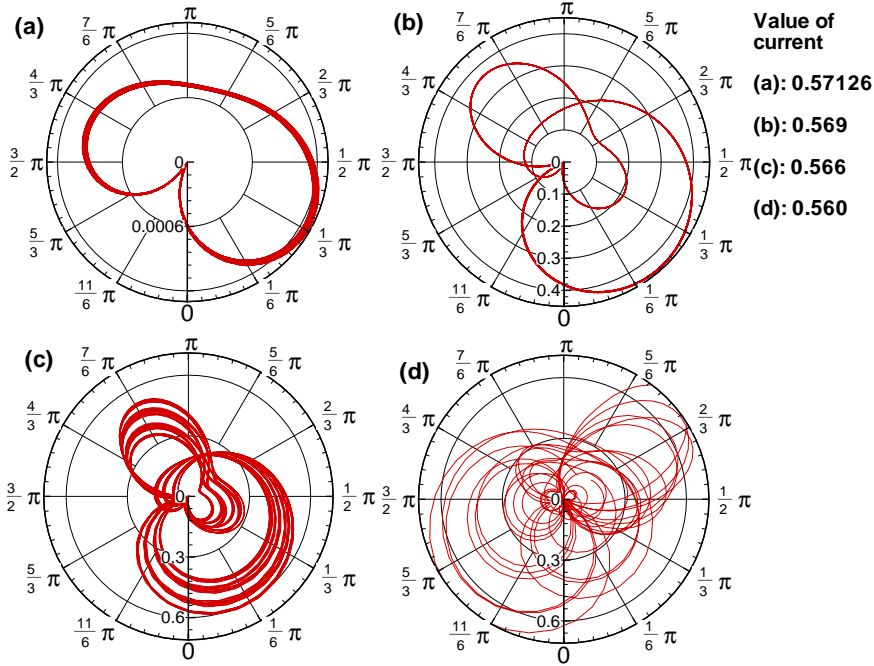}
\caption{\label{polar}(Color online) The phase-charge diagram (the charge on the first superconducting layer
versus the phase difference in the first junction) at different value of bias current.  The radial axis is a
charge, the azimuthal axis is a phase difference.}
\end{figure}

Fig.~\ref{polar} shows such phase-charge diagrams: the variation in time of the modular of charge on the first
superconducting layer $Q_1$  and phase difference $\varphi_1$ in the first Josephson junction  of the stack. In
these figures the radial axis shows the charge $Q_1$ and azimuthal is the phase $\varphi_1$.

\begin{figure}[h]
\centering
\includegraphics[height=90mm]{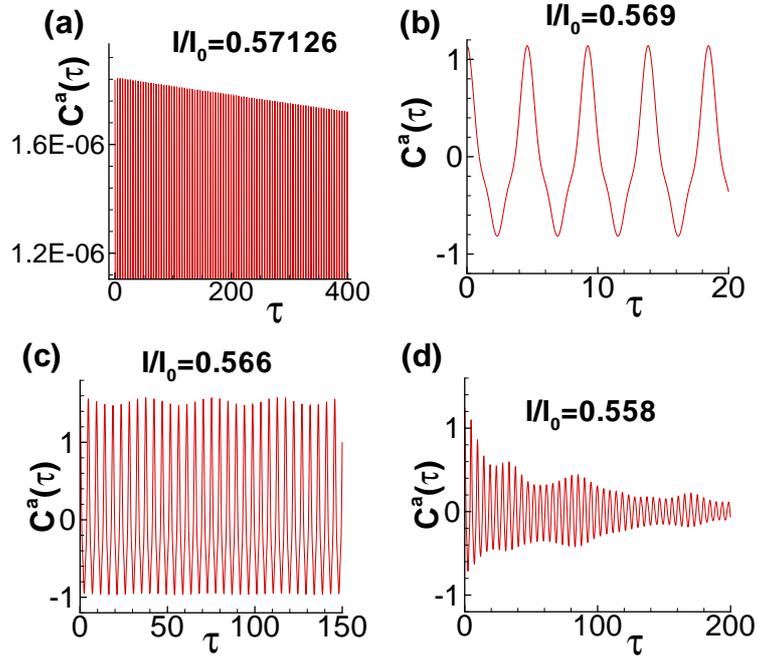}
\caption{\label{Ca}(Color online) The charge autocorrelation function $C^a$ for the first layer at different
values of bias current in the breakpoint region.}
\end{figure}

The beginning part of BPR corresponds to the onset of the parametric resonance: charge on layers is growing.  In
this region, particularly,  $I=0.57126$ (Fig.~\ref{polar}a) the Josephson frequency is still high than LPW
frequency, so the trajectory in phase-charge space has some small thickness. At $I=0.569$ (see
Fig.~\ref{polar}b) the conditions $\omega_J=2\omega_{LPW}$ is practically ideally fulfilled, charge on layers
has reached its maximum amplitude, and thickness of trajectory in phase-charge space is just thickness of line.
Results of FFT analysis (not presented here) show that at this value of current we observe the peaks
corresponding to the LPW only and Josephson frequency does not manifests itself.  The system follows  this
trajectory all time. As we show below, the  results of analysis of autocorrelation function, presented in
Fig.~\ref{Ca} confirm this interpretation \cite{shk-prb09}.

At $I=0.566$ (Fig.~\ref{polar}c)  the trajectory in phase-charge space is wide again. The resonance is passed:
$\omega_J>2\omega_{LPW}$. Here the amplitude of charge on the layers is merely constant but there is a beating
of LPW in the stack. A new large  period of system appears \cite{sms-prb08},  but the system after some cycles
repeats the same trajectory. Analysis of charge correlations shows that the behavior of system is regular, while
Fig.~\ref{polar}d corresponds to a chaotic behavior of the system.

There is a correspondence between phase-charge diagrams and autocorrelation functions, presented in
Fig.\ref{Ca}. The smallest value of  $\tau=\tau_0$ with $C^{a}(\tau=\tau_0)=C^{a}(\tau=0)$  is a period of
system. If the autocorrelation function has not such points, then the trajectory of system in phase-charge
diagram is open. There is such $\tau=\tau_0$ in the cases presented in Figs.~\ref{Ca}b and ~\ref{Ca}c which
satisfies the mentioned condition and that's why  in Figs \ref{polar}b and \ref{polar}c the trajectories in
phase space are closed. But in Fig. \ref{Ca}d the autocorrelation function doesn't show such $\tau_0$ and hence
Fig. \ref{polar}d show open trajectories. Behavior of the system in this case is chaotic. As we mentioned above
the thickness of trajectory in Fig.\ref{Ca}a is related to the growing charge on the superconducting layers and
this is a reason of the amplitude's  decreasing of autocorrelation function with time. Note that to show the
very small decreasing of autocorrelation we zoomed the region in vertical axis of Fig.\ref{Ca}a. The value
$\tau$ is introduced in the definition of autocorrelation function
\begin{eqnarray}
C^{a}_l({\tau})=\lim_{(Tf-Ti)\rightarrow\infty}\frac{1}{Tm-Ti}\int_{Ti}^{T_m}{Q_l(t-\tau)Q_l(t)dt} \label{Ca-eq}
\end{eqnarray}
$C^a(\tau=0)$ is a maximum and its value is $C^{a}_l(\tau=0)=<Q^{2}_{l}(t)>$.  At $I=0.566$ (Fig.~\ref{polar}c)
the trajectory in phase-charge space is wide again. The resonance is passed: $\omega_J>2\omega_{LPW}$. Here the
amplitude of charge on the layers is merely constant.

\section{Summary}
We found that the  phase-charge diagram reflects the dynamical behavior of IJJ. It was shown that the results of
the phase-charge diagram analysis  are in agreement with the results of the autocorrelation function analysis.
We demonstrated that using the phase-charge diagrams (phase portraits) give us a powerful method for
investigation of the fine structure in the CVC.

\ack We thank M.R. Kolahchi for his fruitful discussions.

\end{document}